\documentclass[12pt]{article}
\usepackage{amsmath,amssymb,epsfig}

\newcommand{\Z}{{\bf Z}}

\renewcommand{\thefootnote}{\fnsymbol{footnote}}

\begin{document}

\title{
\begin{flushright}
\ \\*[-80pt]
\begin{minipage}{0.25\linewidth}
\normalsize
KANAZAWA-07-05\\
KUNS-2073 \\*[50pt]
\end{minipage}
\end{flushright}
{\Large \bf
Discrete R-symmetry anomalies \\ in heterotic orbifold models
\\*[20pt]}}

\author{Takeshi~Araki$^{1}$, \
Kang-Sin~Choi$^{2}$, \
Tatsuo~Kobayashi$^{3}$, \\ 
Jisuke~Kubo$^{1}$
 \ and \ 
Hiroshi~Ohki$^{4}$
\\*[20pt]
$^1${\it \normalsize 
Institute for Theoretical Physics, Kanazawa University, Kanazawa
920-1192, Japan}\\
$^2${\it \normalsize
Physikalisches Institut, Universit\" at Bonn,
Nu\ss alle 12, D-53115 Bonn, Germany} \\
$^3${\it \normalsize
Department of Physics, Kyoto University,
Kyoto 606-8502, Japan} \\
$^4${\it \normalsize
Department of Physics, Kyoto University,
Kyoto 606-8501, Japan} 
\\*[50pt]}

\date{
\centerline{\small \bf Abstract}
\begin{minipage}{0.9\linewidth}
\medskip
\medskip
\small
Anomalies of discrete R-symmetries
appearing in heterotic orbifold models are 
studied. We find that
the mixed anomalies for different gauge groups 
satisfy the universal Green-Schwarz (GS) condition,
indicating  that these anomalies 
are canceled by the GS mechanism.
An exact relation between  
the anomaly coefficients
of the discrete R-symmetries 
and one-loop beta-function coefficients is obtained.
We also find that the discrete R-symmetries 
have a good chance to be unbroken down to
the supersymmetry breaking scale.
Even below this scale a $Z_2$ subgroup 
is  unbroken, and it may be an origin of 
the R-parity of the minimal supersymmetric standard model.
Relations between the
R-symmetry anomalies and T-duality anomalies 
are  also investigated.
\end{minipage}
}

\begin{titlepage}
\maketitle
\thispagestyle{empty}
\end{titlepage}

\renewcommand{\thefootnote}{\arabic{footnote}}
\setcounter{footnote}{0}

\section{Introduction}

Discrete symmetries play an important role in 
model building of particle physics.
For example, abelian and non-abelian discrete flavor symmetries 
are useful to derive realistic quark/lepton masses and their 
mixing \cite{Altarelli:2007cd}.
Discrete non-abelian flavor symmetries can also be used to 
suppress flavor changing neutral current processes in 
supersymmetric models \cite{dbkaplan,babu}.
Furthermore, discrete symmetries can be introduced to forbid unfavorable couplings 
such as those leading to fast proton decay \cite{murayama,kakizaki}.

Superstring theory is a promising candidate for 
unified theory including gravity and
 may provide with an origin of 
such discrete symmetries \cite{Dine:1988kq}.
It is widely assumed that
superstring theory leads to anomaly-free effective theories.
In fact the anomalous $U(1)$ symmetries
are restored by the Green-Schwarz (GS) 
mechanism \cite{Green:1984sg,Witten:1984dg,Ibanez:1998qp}.
For this mechanism to work,
the mixed anomalies between the anomalous $U(1)$ and other continuous 
gauge symmetries have to satisfy a certain set of conditions,
the GS conditions,
at the field theory level.
In particular, in heterotic string theory the mixed anomalies between 
the anomalous $U(1)$ symmetries and other continuous gauge 
symmetries must be universal for different gauge 
groups up to their Kac-Moody 
levels \cite{Schellekens:1986xh,Kobayashi:1996pb}.
A well-known discrete symmetry in heterotic string theory is
 T-duality symmetry, and 
its effective theory has T-duality anomalies \cite{Derendinger:1991hq}.
It has been shown that
the mixed anomalies between T-duality symmetry and 
continuous gauge symmetries are universal except 
for the sector containing an $N=2$ subsector
 and are exactly canceled by the GS mechanism  \cite{Ibanez:1992hc}.
That has phenomenologically interesting 
consequences which  have been  studied in early 90's
\cite{Ibanez:1992hc,Ibanez:1991zv,Kawabe:1993pz}.

Heterotic orbifold construction is one of interesting 
4D string models \cite{Dixon,IMNQ}.
(See also for resent works Ref.~\cite{Kobayashi:2004ud,Forste:2004ie} 
and for review \cite{Choi-Km}.)
Geometrical structures of their compact spaces are simple compared with 
other types of 4D string model constructions.
Phenomenological aspects in effective theory are related 
with geometrical aspects of orbifolds.
Discrete symmetries which may be used as  non-abelian flavor symmetries 
and also certain discrete $R$-symmetries originate from the geometrical 
structure of 
orbifolds \cite{Dine:1988kq,Ibanez:1992uh,Kobayashi:2004ud,Kobayashi:2006wq}.
In this paper
we consider discrete R-symmetries.
Stringy-originated discrete symmetries are strongly constrained
 due to stringy consistency, and 
 it is phenomenologically  and theoretically
 important to study  anomalies of discrete symmetries,
 as it is pointed out in \cite{Ibanez:1991hv}
and the example of T-duality shows.
We shall investigate the mixed anomalies between the discrete R-symmetries 
and the continuous gauge symmetries in concrete orbifold models. 
We will also study relations between the discrete $R$-anomalies, 
one-loop beta-function coefficients (scale anomalies) and 
T-duality anomalies.

This paper is organized as follows.
In section 2, we give a brief review on heterotic 
orbifold models to fix our notation.
In section 3, we define discrete $R$-charges, which is
one of our main interests. 
In section 4, we calculate the mixed anomalies between 
the discrete $R$-symmetries and the continuous gauge symmetries
in concrete models.
We also study the relations of R-anomalies with 
one-loop beta-function coefficients and T-duality anomalies.
In section 5, we discuss phenomenological implications of 
our results.
Section 6 is devoted to conclusion and discussion.

\section{Heterotic orbifold models}

Here we review briefly heterotic orbifold models.
First we give a review on $\Z_N$ orbifold models, and next 
explain $\Z_N \times \Z_M$ orbifold models.
Heterotic string theory consists of 
10D right-moving superstrings and 26D left-moving 
bosonic strings.
Their common 10 dimensions correspond to our 4D space-time 
and 6D compact space.
The other 16D left-moving bosonic strings correspond to 
a gauge part.
Here, we consider the $E_8 \times E_8$ heterotic string theory, 
where momenta of 16D left-moving bosonic strings span 
$E_8 \times E_8$ root lattice.
The following discussions are also applicable to 
$SO(32)$ heterotic string theory.

In orbifold models, the 6D compact space is chosen to be   
6D orbifold.
A 6D orbifold is a division of 6D torus $T^6$ by a twist $\theta$, 
while the torus $T^6$ is obtained as $R^6/\Lambda^6$, 
where $\Lambda^6$ is 6D lattice.
Eigenvalues of the twist $\theta$ are denoted as 
$e^{2\pi i v_1}, e^{2\pi i v_2}$ 
and $e^{2\pi i v_3}$ in the complex basis $Z_i$ ($i=1,2,3$).
To preserve 4D N=1 supersymmetry (SUSY), they must satisfy 
the following condition,
\begin{equation}
v_1+v_2+v_3= {~~\rm integer}.
\end{equation}
When one of $v_i$ is integer, N=2 SUSY is preserved.
In the case with $v_i\neq {\rm integer}$, only N=1 SUSY 
is preserved.
Such $\Z_N$ orbifolds are classified into 
$\Z_3$, $\Z_4$, $\Z_6$-I, $\Z_6$-II, $\Z_7$,  $\Z_8$-I, $\Z_8$-II, 
$\Z_{12}$-I and $\Z_{12}$-II, and their twists are explicitly shown 
in Table 1 and Table 2.

\begin{table}[t]
\begin{center}
\small
\begin{tabular}{|c|c|c|c|c|c|}
\hline
 & $\Z_3$ & $\Z_4$ & $\Z_6$-I & $\Z_6$-II & $\Z_7$  \\
$v_i$ & $(1,1,-2)/3$ & $(1,1,-2)/4$ & $(1,1,-2)/6$ & $(1,2,-3)/6$ &
$(1,2,-3)/7$  \\ \hline \hline
$T_1$ & $(1,1,1)/3$ & $(1,1,2)/4$ & $(1,1,4)/6$ & $(1,2,3)/6$  &
$(1,2,4)/7$ \\
$T_2$ & --- & $(2,2,0)/4$ & $(2,2,2)/6$ & $(2,4,0)/6$ & $(2,4,1)/7$ \\
$T_3$ & --- & --- & $(3,3,0)/6$ &  $(3,0,3)/6$ &  --- \\
$T_4$ & --- & --- & --- & $(4,2,0)/6$ & $(4,1,2)/7$ \\ \hline
\end{tabular}
\end{center}
\caption{$H$-momenta for $\Z_3$, $\Z_4$, $\Z_6$-I, $\Z_6$-II and
$\Z_7$ orbifolds} \label{tab:H-momenta-1}
\end{table}

\begin{table}[t]
\begin{center}
\small
\begin{tabular}{|c|c|c|c|c|}
\hline
 &  $\Z_8$-I & $\Z_8$-II & $\Z_{12}$-I & $\Z_{12}$-II  \\
$v_i$ &  $(1,2,-3)/8$ & $(1,3,-4)/8$ & $(1,4,-5)/12 $ &
$(1,5,-6)/12$ \\  \hline \hline
$T_1$ & $(1,2,5)/8$ & $(1,3,4)/8$ & $(1,4,7)/12$ & $(1,5,6)/12$ \\
$T_2$ & $(2,4,2)/8$ & $(2,6,0)/8$ & $(2,8,2)/12$ & $(2,10,0)/12$ \\
$T_3$ & ---         & $(3,1,4)/8$ & $(3,0,9)/12$ & $(3,3,6)/12$ \\
$T_4$ & $(4,0,4)/8$ & $(4,4,0)/8$ & $(4,4,4)/12$ & $(4,8,0)/12$ \\
$T_5$ & $(5,2,1)/8$ & ---         & ---          & $(5,1,6)/12$ \\
$T_6$ & ---         & ---         & $(6,0,6)/12$ & $(6,6,0)/12$ \\
$T_7$ & ---         & ---         & $(7,4,1)/12$ & ---          \\
$T_8$ & ---         & ---         & ---          & ---          \\
$T_9$ & ---         & ---         & $(9,0,3)/12$ & ---          \\
$T_{10}$ & ---      & ---         & ---          & $(10,2,0)/12$ \\
\hline
\end{tabular}
\end{center}
\caption{$H$-momenta for $\Z_8$-I, $\Z_8$-II, $\Z_{12}$-I and
$\Z_{12}$-II orbifolds} \label{tab:H-momenta-2}
\end{table}

On the orbifold, closed string satisfies the following boundary 
condition,
\begin{equation}
X(\sigma = \pi) = \theta^k X (\sigma = 0) +V,
\end{equation}
where $V$ is a shift vector on the 6D lattice $\Lambda^6$.
The complex basis of $X$ corresponds to $Z_i$.
The $\theta^k$-twisted sector is denoted by $T_k$, while 
the sector with $k=0$ is the so-called untwisted sector.

It is convenient to bosonize right-moving fermionic strings.
Here we write such bosonized fields by $H^t$ ($t=1,\cdots,5$).
Their momenta $p_t$ are quantized and span the SO(10) weight lattice.
Space-time bosons correspond to SO(10) vector momenta, 
and space-time fermions correspond to SO(10) spinor momenta.
The 6D compact part, i.e. the SO(6) part, $p_i$ ($i=1,2,3$) 
is relevant to our study.
All of $\Z_N$ orbifold models have three untwisted 
sectors, $U_1$, $U_2$ and $U_3$, and their massless bosonic modes 
have the following SO(6) momenta,
\begin{equation}
U_1:(1,0,0), \qquad U_2:(0,1,0), \qquad U_3:(0,0,1).
\label{H-momenta-U}
\end{equation}
On the other hand, the twisted sector $T_k$ has 
shifted $SO(6)$ momenta, $r_i=p_i+kv_i$.
Table 1 and Table 2 show explicitly $H$-momenta $r_i$ of 
massless bosonic states.
That implies their $SO(6)$ $H$-momenta are obtained as 
\begin{equation}
r_i = |kv_i|-{\rm Int}[|kv_i|],
\label{H-momentum:Zn}
\end{equation}
where ${\rm Int}[a]$ denotes an integer part of fractional number $a$.
This relation is not available for the untwisted sectors, 
and $r_i$ is obtained as Eq.~(\ref{H-momenta-U}).

The gauge sector can also be broken and 
gauge groups smaller than $E_8 \times E_8$ are obtained.
Matter fields have some representations under such 
unbroken gauge symmetries.

Massless modes for 4D space-time bosons correspond to 
the following vertex operator \cite{Friedan:1985ge,Dixon:1986qv},
\begin{equation}
V_{-1} = e^{-\phi}\prod_{i=1}^3(\partial Z_i)^{{\cal N}_i} (\partial \bar
Z_i)^{\bar {\cal N}_i}e^{ir_tH^t}e^{iP^IX^I}e^{ikX}
\sigma_k,
\end{equation}
in the $(-1)$-picture, where $\phi$ is the bosonized ghost, 
$kX$ corresponds to the 4D part and $P^IX^I$ corresponds 
to the gauge part.
Oscillators of the left-mover are denoted by 
$\partial Z_i$ and $\partial \bar Z_i$, and 
${\cal N}_i$ and $\bar {\cal N}_i$ are oscillator numbers, which are included 
in these massless modes.
In addition, $\sigma_k$ denotes the twist field for 
the $T_k$ sector.
Similarly, we can write the vertex operator for 4D space-time 
massless fermions as 
\begin{equation}
V_{-\frac12} = e^{-\frac12 \phi}\prod_{i=1}^3(\partial Z_i)^{N_i}
(\partial \bar Z_i)^{\bar N_i}e^{ir_t^{(f)}H_t}e^{iP^IX^I}e^{ikX}
\sigma_{k},
\end{equation}
in the $(-1/2)$-picture.
The $H$-momenta for space-time fermion and boson, $r_i^{(f)}$ and
$r_i$ in the same supersymmetric multiplet are
related each other as 
\begin{equation}
r_i = r_i^{(f)} + (1,1,1)/2.
\end{equation}

We need vertex operators $V_0$ with the 0-picture when we compute 
generic n-point couplings.
We can obtain such vertex operators $V_0$ by operating 
the picture changing operator, $Q$, on $V_{-1}$, 
\cite{Friedan:1985ge},
\begin{equation}
Q=e^\phi (e^{-2 \pi i r^v_i H_i}\bar \partial Z_i + e^{2 \pi i r^v_i
H_i}\bar \partial \bar Z_i), \label{p-change}
\end{equation}
where $r^v_1=(1,0,0)$, $r^v_2=(0,1,0)$ and $r^v_3=(0,0,1)$.

Next we briefly review on $\Z_N \times \Z_M$ orbifold 
models \cite{Font:1988mk}.
In $\Z_N \times \Z_M$ orbifold models, we introduce two independent 
twists $\theta$ and $\omega$, whose twists are represented by 
$e^{2\pi i v^1_i}$ and $e^{2\pi i v^2_i}$, respectively 
in the complex basis.
Two twists are chosen such that 
each of them breaks 4D N=4 SUSY to 4D N=2 SUSY and 
their combination preserves only N=1 SUSY.
Thus, eigenvalues $v^1_i$ and $v^2_i$ are chosen as 
\begin{equation}
v^1_i=(v^1,-v^1,0), \qquad v^2_i=(0,v^2,-v^2),
\end{equation}
where $v^1,v^2 \neq {\rm integer}$.
In general, $\Z_N \times \Z_M$ orbifold models 
have three untwisted sectors, $U_1$, $U_2$ and $U_3$, 
and their massless bosonic modes have the same $SO(6)$ $H$-momenta $r_i$ as 
Eq.~(\ref{H-momenta-U}).
In addition, there are $\theta^k \omega^\ell$-twisted sectors, 
and their $SO(6)$ $H$-momenta are obtained as 
\begin{equation}
r_i = |kv^1_i|+|\ell v^2_i| - {\rm Int}[|kv^1_i|+|\ell v^2_i|].
\label{H-momentum:ZnZm}
\end{equation}
Vertex operators are also constructed in a similar way.
Recently, non-factorizable $\Z_N \times \Z_M$ orbifold 
models have been studied \cite{Faraggi:2006bs}.
The above aspects are the same for such non-factorizable models.

\section{Discrete R-symmetries}

Here we define R-charges.
We consider n-point couplings including two fermions.
Such couplings are computed by the following 
n-point correlation function of vertex operators,
\begin{equation}
\langle V_{-1}V_{-1/2}V_{-1/2}V_0\cdots V_0 \rangle .
\end{equation}
They must have the total ghost charge $-2$, because the background 
has the ghost number 2.
When this n-point correlation function does not vanish, 
its corresponding n-point coupling in effective theory is 
allowed.
That is, selection rules for allowed n-point correlation functions 
in string theory correspond to symmetries in effective theory.

The vertex operator consists of several parts, the 
4D part $e^{kX}$, the gauge part $e^{iPX}$, 
the 6D twist field $\sigma_k$, the 6D left-moving oscillators 
$\partial Z_i$ and the bosonized fermion $e^{irH}$.
Each part has its own selection rule for allowed couplings.
For the 4D part and the gauge part, 
the total 4D momentum $\sum k$ and the total momentum 
of the gauge part $\sum P$ should be conserved.
The latter is nothing but the requirement of gauge invariance.
The selection rule for 6D twist fields $\sigma_k$ is controlled by 
the space group selection rule 
\cite{Dixon:1986qv,Kobayashi:1991rp}.

Similarly, the total $H$-momenta can be conserved 
\begin{equation}
\sum r_i =1.
\end{equation}
Here we take a summation over the $H$-momenta for 
scalar components, using the fact that 
the $H$-momentum of fermion component differs by $-1/2$.
Another important symmetry is the twist symmetry of
oscillators.
We consider the following twist of oscillators,
\begin{eqnarray}
& & \partial Z_i \rightarrow e^{2 \pi i v_i}\partial Z_i, \qquad
\partial \bar Z_i \rightarrow e^{-2 \pi i v_i}\partial \bar Z_i, 
\nonumber \\
& & \bar \partial Z_i \rightarrow e^{2 \pi i v_i}\bar \partial Z_i,
\qquad \bar \partial \bar Z_i \rightarrow e^{-2 \pi i v_i}\bar
\partial \bar Z_i.
\end{eqnarray}
Allowed couplings may be invariant under the above $Z_N$ twist.

Indeed, for 3-point couplings corresponding to 
$\langle V_{-1}V_{-1/2}V_{-1/2}\rangle$, we can require 
$H$-momentum conservation and $Z_N$ twist invariance of oscillators 
independently.
However, we have to compute generic n-point couplings 
through picture changing, and the picture changing operator $Q$ 
includes non-vanishing $H$-momenta and right-moving oscillators 
$\bar \partial
Z_i$ and
 $\bar \partial \bar Z_i$. 
Consequently, the definition of the H-momentum  
of each vertex operator depends on the choice of
the picture and so its physical meaning remains somewhat obscure.
We therefore use a picture independent quantity as 
follows,
\begin{equation}
R_i \equiv r_i + {\cal N}_i - \bar {\cal N}_i,
\end{equation}
which can be interpreted as an R-charge  \cite{Kobayashi:2004ud}.
This R-symmetry is a discrete 
surviving symmetry of the continuous $SU(3)~(\subset SU(4))$ R-symmetry 
under orbifolding. 
Here we do not
distinguish oscillator numbers for the left-movers and right-movers,
because they have the same phase under $Z_N$ twist. Indeed,
physical states with $-1$ picture have vanishing oscillator number
for the right-movers, while the oscillator number for the
left-movers can be non-vanishing. Thus, hereafter   ${\cal N}_i$ and
$\bar {\cal N}_i$ denote the oscillator number for the left-movers,
because we study the physical states with $-1$ picture from now. For
simplicity, we use the notation $\Delta {\cal N}_i = {\cal N}_i -
\bar {\cal N}_i$. 
Now, we can write the selection rule due to $R$-symmetry as 
\begin{equation}
\sum R_i = 1 \quad {\rm mod} \quad N_i,
\end{equation}
where $N_i$ is the minimum integer satisfying $N_i = 1/\hat v_i$, 
where $\hat v_i= v_i + m$ with any integer $m$. 
For example, for $Z_6$-II orbifold, we have $v_i=(1,2,-3)/6$, and
$N_i=(6,3,2)$.
Thus, these are discrete symmetries.
Note that the above summation is taken over scalar components.

Discrete R symmetry itself is defined as the following 
transformation,
\begin{equation}
\vert R_i \rangle \rightarrow e^{2\pi i v_i R_i} \vert R_i \rangle,
\label{eq:R-trans}
\end{equation}
for states with discrete $R$-charges, which are defined 
mod $N_i$.
For later convenience, we show discrete $R$-charges for 
fermions in Table~\ref{tab:R}.
As shown there, gaugino fields always have 
$R$-charge $(1/2,1/2,1/2)$.

\begin{table}[t]
\begin{center}
\small
\begin{tabular}{|c|c|}  \hline
 & $R_i$   \\ \hline
gaugino & $(1/2,1/2,1/2)$ \\
$U_1$   & $(1/2,-1/2,-1/2)$ \\
$U_2$   & $(-1/2,1/2,-1/2)$  \\
$U_3$ & $(-1/2,-1/2,1/2)$  \\
$T_k$ & $kv_i - {\rm Int}[kv_i]-1/2+\Delta {\cal N}_i$ \\ \hline
\end{tabular}
\end{center}
\caption{Discrete $R$-charges of fermions in 
$\Z_N$ orbifold models} \label{tab:R}
\end{table}

\section{Anomalies of R-symmetry}

\subsection{Discrete R anomalies}

Let us study anomalies of discrete R-symmetry.
Under the R-transformation like Eq.~(\ref{eq:R-trans}), the 
path integral measure of fermion fields is not 
invariant, but changes as 
\begin{equation}
{\cal D}\psi {\cal D}\psi^\dagger \rightarrow 
{\cal D}\psi {\cal D}\psi^\dagger exp \left[-2\pi i v_i \sum_{G_a} A^{R_i}_{G_a}
\int d^4x \frac{1}{16\pi^2}F^{(G_a)}_{\mu \nu}\tilde F^{{(G_a)} \mu \nu}\right],
\end{equation}
where $\tilde F^{{(G_a)} \mu \nu}=\frac{1}{2}\varepsilon^{\mu \nu \rho
  \sigma} F^{(G_a)}_{\rho \sigma}$.
The anomaly coefficients $A^{R_i}_{G_a}$ are obtained as 
\begin{equation}
A_{G_a}^{R_i}=\sum R_i T({\bf R}_{G_a}),
\end{equation}
where $T({\bf R}_{G_a})$ is the Dynkin index for ${\bf R}_{G_a}$
representation under $G_a$.
The winding number of the gauge field configuration, 
i.e., the Pontryagin index, 
\begin{equation}
\nu \equiv \frac{T({\bf R}^{(f)}_{G_a})}{16\pi^2}\int  d^4x F^{(G_a)}_{\mu \nu}\tilde F^{{(G_a)} \mu \nu},
\end{equation}
is integer, where $T({\bf R}^{(f)}_{G_a})$ denotes 
the Dynkin index of a fundamental representation of $G_a$.
Thus, the anomaly coefficients $A^{R_i}_{G_a}$ 
are defined modulo $N_iT({\bf R}^{(f)}_{G_a})$.

By use of our discrete $R$ charge, the anomaly coefficients are written as 
\begin{equation}
A_{G_a}^{R_i}= \frac{1}{2} C_2(G_a) + 
\sum_{\rm matter}  (r^{}_i-\frac{1}{2} +\Delta {\cal N}_i) T({\bf R}_{G_a}),
\end{equation}
where $C_2(G_a)$ is quadratic Casimir.
Note that $r_i$ denotes the SO(6) shifted momentum for 
bosonic states.
The first term in the right hand side is a contribution from 
gaugino fields and the other is the contribution from matter fields.

If these anomalies are canceled by the 
Green-Schwarz mechanism, these mixed anomalies must 
satisfy the following condition,
\begin{equation}
\frac{A_{G_a}^{R_{i}}}{k_a}= \frac{A_{G_b}^{R_i}}{k_b},
\label{GS-R}
\end{equation}
for different gauge groups, $G_a$ and $G_b$, where 
$k_a$ and $k_b$ are Kac-Moody levels.
In the simple orbifold construction, we have the Kac-Moody level 
$k_a=1$ for non-abelian gauge groups.
Note again that anomalies are defined modulo
$N_iT({\bf R}^{(f)}_{G_a})$.
The above GS condition has its meaning mod
$N_iT({\bf R}^{(f)}_{G_a})/k_a$.

As illustrating examples, let us study explicitly one $Z_3$ model 
and one $Z_4$ model.
Their gauge groups and massless spectra are shown 
in Table~\ref{tab:Z3} and Table~\ref{tab:Z4}.\footnote{
See for explicit massless spectra Ref.~\cite{Katsuki:1989cs}, 
where a typographical error is included in the $U_3$ sector 
of the $Z_4$ orbifold model.
It is corrected in Table~\ref{tab:Z4}.}
First, we study R-anomalies in the $Z_3$ orbifold model.
Since $v_i=(1,1,-2)/3$, we have $N_i=3$.
For both $E_6$, mixed R-anomalies are computed as 
\begin{equation}
A^{R_{i}}_{E_6}= \frac{3}{2}+9n^i_{E_6},
\end{equation}
where $n^i_{E_6}$ is integer.
The second term in the right hand side appears because 
anomalies are defined modulo $N_iT(27)$ with 
$N_i=3$ and $T(27)=3$ for $E_6$.
Similarly, mixed R-anomalies for $SU(3)$ are computed as 
\begin{equation}
A^{R_i}_{SU(3)}=-12 +\frac{3}{2}n^i_{SU(3)},
\end{equation}
where $n^i_{SU(3)}$ is integer.
The second term in the right hand side appears through 
$N_iT(3)$ with $N_i=3$ and $T(3)=1/2$ for $SU(3)$.
Thus, in this model, mixed R-anomalies satisfy 
\begin{equation}
A^{R_i}_{E_6}=A^{R_i}_{SU(3)} \qquad ({\rm mod}~~3/2).
\end{equation}

\begin{table}[t]
\begin{center}
\small
\begin{tabular}{|c|c|}  \hline
gauge group & $E_6 \times SU(3) \times E_6 \times SU(3)$ \\ \hline \hline
sector & massless spectrum   \\ \hline
$U_1$   & (27,3;1,1)+ (1,1;27,3)\\
$U_2$   & (27,3;1,1)+ (1,1;27,3)  \\
$U_3$ & (27,3;1,1)+ (1,1;27,3)  \\  
$T_1$ & $27(1,\bar 3;1,\bar3)$ \\ \hline
\end{tabular}
\end{center}
\caption{Massless spectrum in a 
$\Z_3$ orbifold model} \label{tab:Z3}
\end{table}

\begin{table}[t]
\begin{center}
\small
\begin{tabular}{|c|c|}  \hline
gauge group & $SO(10)  \times SU(4) \times SO(12) \times SU(2)
\times U(1)$ \\ \hline \hline
sector & massless spectrum   \\ \hline
$U_1$   & $(16_c,4;1,1)+ (1,1;32_c,1)+(1,1;12_v,2)$ \\
$U_2$   & $(16_c,4;1,1)+  (1,1;32_c,1)+(1,1;12_v,2)$           \\
$U_3$ & $(10_v,6;1,1)+ (1,1;32_c,2) +2(1,1,;1,1)$  \\
$T_1$ & $16(1,4;1,2)$ \\ 
$T_2$ & $16(10_v,1;1,1)+16(1,6;1,1)$ \\ \hline
\end{tabular}
\end{center}
\caption{Massless spectrum in a 
$\Z_4$ orbifold model} \label{tab:Z4}
\end{table}

Next, we study R-anomalies in the $Z_4$ orbifold model 
with the gauge group 
$SO(10)\times SU(4) \times SO(12) \times SU(2) \times U(1)$.
Since the $Z_4$ orbifold has $v_i=(1,1,-2)/4$, 
we have $N_i=(4,4,2)$.
Mixed anomalies between $R_{1,2}$ and $SO(10)$ are 
computed as 
\begin{equation}
A^{R_{1,2}}_{SO(10)} = 1 + 4 n^{1,2}_{SO(10)},
\end{equation}
with integer $n^{1,2}_{SO(10)}$, 
where the second term appears through $N_iT({\bf R}_a)$ 
with $N_i=4$ and $T(10)=1$ for $SO(10)$.
Similarly, mixed anomalies between $R_3$ and $SO(10)$ 
is computed as 
\begin{equation}
A^{R_{3}}_{SO(10)} = -9 + 2 n^{3}_{SO(10)},
\end{equation}
with integer $n^{3}_{SO(10)}$.
Furthermore, mixed R-anomalies for other non-abelian groups 
are obtained as 
\begin{eqnarray}
 & & A^{R_{1,2}}_{SU(4)} = -7 + 2 n^{1,2}_{SU(4)},  \qquad 
A^{R_{3}}_{SU(4)} = -9 + n^{3}_{SU(4)},  \nonumber \\
 & & A^{R_{1,2}}_{SO(12)} = 1 + 4 n^{1,2}_{SO(12)},  \qquad 
A^{R_{3}}_{SO(12)} = 3 + 2n^{3}_{SO(12)},   \\
& & A^{R_{1,2}}_{SU(2)} = -15 + 2 n^{1,2}_{SU(2)},  \qquad 
A^{R_{3}}_{SU(2)} = 3 + n^{3}_{SU(2)},  
\nonumber 
\end{eqnarray}
with integer $n^{i}_{G_a}$, where the second terms 
appear through $N_iT({\bf R}_a)$ with $N_i=(4,4,2)$, and 
$T(12)=1$ for $SO(12)$, $T(4)=1/2$ for $SU(4)$ and 
$T(2)=1/2$ for $SU(2)$.
These anomalies satisfy the GS condition,
\begin{eqnarray}
 & & A^{R_{1,2}}_{SO(10)}=A^{R_{1,2}}_{SU(4)} =  A^{R_{1,2}}_{SO(12)} = 
A^{R_{1,2}}_{SU(2)} \qquad ({\rm mod}~~2), \nonumber \\
 & & A^{R_{3}}_{SO(10)}=A^{R_{3}}_{SU(4)} =  A^{R_{3}}_{SO(12)} = 
A^{R_{3}}_{SU(2)} \qquad ({\rm mod}~~1).
\end{eqnarray}

\subsection{Relation with beta-function}

Here we study the relation between discrete R anomalies and 
one-loop beta-functions.
We find 
\begin{equation}
\sum_{i=1,2,3}r_i=1,
\end{equation}
{}from Eqs.~(\ref{H-momentum:Zn}) and (\ref{H-momentum:ZnZm}) 
as well as Table~\ref{tab:H-momenta-1} and 
Table~\ref{tab:H-momenta-2}.
By using this, we can write the sum of R-anomalies as 
\begin{eqnarray}
A^R_{G_a} &=& \sum_{i=1,2,3}A^{R_i}_{G_a} \nonumber \\
  &=& \frac{3}{2}C_2(G_a) + \sum_{\rm matter}  T({\bf R}_{G_a})(-\frac{1}{2}+ 
\sum_i\Delta {\cal N}_i).
\end{eqnarray}
Thus, when $\sum_i\Delta {\cal N}_i=0$, the total anomaly 
$A^R_{G_a}$ is proportional to the one-loop beta-function 
coefficient, i.e. the scale anomaly, $b_{G_a}$,
\begin{equation}
b_{G_a} = 3 C_2(G_a) - \sum_{\rm matter} T({\bf R}_{G_a}).
\end{equation}
When we use the definition of R charge $\tilde R_i = 2 R_i$, 
we would have $A^{\tilde R}_{G_a} = b_{G_a}$.
It is not accidental that 
$A^R_{G_a}$ is proportional to $b_{G_a}$
\cite{jones,piguet1}.
The sum of the R-charges $\sum_{i=1,2,3}R_i$
  of a supermultiplet is
nothing but the R-charge (up to an overall normalization)
associated with the R-current
which is a bosonic component of the supercurrent \cite{ferrara}, 
when the R-charge is universal for all of matter fields, 
i.e. $\sum_i\Delta {\cal N}_i=0$. 
Using the supertrace identity \cite{piguet2} it is
in fact possible to show \cite{piguet1} that
$A^R_{G_a}$ is proportional to $b_{G_a}$ to all orders in perturbation 
theory.

In explicit models, non-abelian groups except $SU(2)$ 
have few massless matter fields with non-vanishing oscillator 
numbers, while massless matter fields with oscillators  
can appear as singlets as well as $SU(2)$ doublets.
Thus, in explicit models the total R-anomaly $A^R_{G_a}$ 
is related with the one-loop beta-function coefficient $b_{G_a}$,
\begin{equation}
2A^R_{G_a} = b_{G_a},
\label{anomR-b}
\end{equation}
modulo $N_iT({\bf R}_a)$ for most of non-abelian groups.
Since the total R-anomalies satisfy the GS condition, 
$A^R_{G_a}=A^R_{G_b}$, the above relation between 
$A^R_{G_a}$ and $b_{G_a}$ leads to 
\begin{equation}
b_{G_a} = b_{G_b},
\label{GS-b}
\end{equation}
modulo $2N_iT({\bf R}_a)$.

For example, the explicit $Z_3$ orbifold model 
and $Z_4$ orbifold model in Table~\ref{tab:Z3} and Table~\ref{tab:Z4} 
have only non-oscillated massless modes except singlets.
The $Z_3$ orbifold model has the following total R-anomalies and 
one-loop beta-function coefficient,
\begin{eqnarray}
 & & A^R_{E_6}=\frac{9}{2}+9n_{E_6}, \qquad 
 b_{E_6} = 9, \nonumber \\
 & & A^R_{SU(3)}=-36 + \frac{3}{2}n_{SU(3)}, \qquad 
b_{SU(3)}=-72.
\end{eqnarray}
Hence, this model satisfy $2A^R_{G_a}=b_{G_a}$ and its 
one-loop beta-function coefficients satisfy
\begin{equation}
b_{E_6}=b_{SU(3)} \qquad ({\rm mod}~~3).
\end{equation}
Similarly, the $Z_4$ orbifold model in Table~\ref{tab:Z4}
has the total R-anomalies and
one-loop beta-function coefficients as,
\begin{eqnarray}
 & & A^R_{SO(10)}= -7 + 2 n_{SO(10)}, \qquad  b_{SO(10)}= -14
 \nonumber \\
 & & A^R_{SU(4)}= -23 + n_{SU(4)}, \qquad b_{SU(4)} = -46   
\nonumber \\
 & & A^R_{SO(12)}=5 + 2n_{SO(10)}, \qquad b_{SO(12)}= 10    \\
 & & A^R_{SU(2)}=-27 + n_{SU(2)}, \qquad b_{SU(2)}= -54.
\nonumber
\end{eqnarray}
Thus, this model also satisfies $2A^R_{G_a}=b_{G_a}$
and its one-loop beta-function coefficients satisfy 
\begin{equation}
b_{SO(10)}=b_{SU(4)}=b_{SO(12)}=b_{SU(2)} \qquad ({\rm mod}~~2).
\end{equation}

\subsection{Relation with T-duality anomaly}

Here we study the relation between R-anomalies and T-duality anomalies.
The relation between R-symmetries and T-duality has also been 
studied in Ref.~\cite{Ibanez:1992uh}.
The T-duality anomalies are obtained 
as \cite{Derendinger:1991hq,Ibanez:1992hc}
\begin{equation}
A^{T_i}_{G_a} = -C_2({G_a}) +\sum_{\rm matter} T({\bf R}_{G_a}) 
(1+2n_i),
\end{equation}
where $n_i$ is the modular weight of matter fields for 
the $i$-th torus.
The modular weight is related with $r_i$ as 
\begin{eqnarray}
n_i &=& -1 {\rm~~for~~} r_i=1,\nonumber \\
&=& 0 {\rm~~for~~} r_i=0,\\
&=&  r_i-1 - \Delta {\cal N}_i {\rm~~for~~} r_i \neq 0,1.
\nonumber
\end{eqnarray}
Note that $n_i = -r_i$ for $r_i=0,1/2,1$.
Thus, in the model, which includes only matter fields with 
 $r_i=0,1/2,1$, the T-duality anomalies and R-anomalies are 
proportional to each other, 
\begin{equation}
A^{T_i}_{G_a} = -2A^{R_i}_{G_a}.
\label{relation:T-R}
\end{equation}
In generic model, such relation is violated, but 
T-duality anomalies and R-anomalies are still related with each other 
as 
\begin{equation}
A^{T_i}_{G_a} = -2A^{R_i}_{G_a} -2 \sum_{r_i \neq 0,1/2,1}
(2  r_i -1).
\end{equation}

T-duality should also satisfy the GS condition,
\begin{equation}
\frac{A^{T_i}_{G_a}}{k_a} =\frac{A^{T_i}_{G_b}}{k_b},
\end{equation}
for the $i$-th torus, which does not include the N=2 subsector.
Thus, the requirement that T-duality anomalies and R-anomalies 
should satisfy the GS condition, leads to a similar condition for 
\begin{equation}
\Delta_a^i= 2 \sum_{r^b_i \neq 0,1/2,1}
(2  r^b_i -1).
\end{equation}

For the $i$-th torus, which includes 
N=2 subsector, T-duality anomalies can be canceled by 
the GS mechanism and T-dependent threshold correction \cite{Dixon:1990pc}.
Thus, for such torus, the T-duality anomalies has no 
constrain from the GS condition.
However, even for such torus, R-anomaly should satisfy 
the GS condition.

For example, the $Z_4$ orbifold model in Table~\ref{tab:Z4} 
has the following T-duality anomalies,
\begin{eqnarray}
& & A^{T_{1,2}}_{SO(10)}=-2, \qquad A^{T_3}_{SO(10)}=18, \nonumber \\
& & A^{T_{1,2}}_{SU(4)}=-2, \qquad A^{T_3}_{SU(4)}=18, \nonumber \\
& & A^{T_{1,2}}_{SO(12)}=-2, \qquad A^{T_3}_{SO(12)}=-6, \\
& & A^{T_{1,2}}_{SU(2)}=-2, \qquad A^{T_3}_{SU(2)}=-6.
\nonumber
\end{eqnarray}
They satisfy the GS condition,
\begin{equation}
A^{T_{1,2}}_{SO(10)}=A^{T_{1,2}}_{SU(4)}=A^{T_{1,2}}_{SO(12)}=
A^{T_{1,2}}_{SU(2)}.
\end{equation}
On the other hand, for the third torus, T-duality anomalies $A^{T_3}_{G_a}$ 
do not satisfy the GS condition, that is, anomalies $A^{T_3}_{G_a}$ 
are not universal, because there is the N=2 subsector 
and one-loop gauge kinetic functions depend on the $T_3$ moduli 
with non-universal coefficients \cite{Dixon:1990pc}.
However, they satisfy 
\begin{eqnarray}
& & A^{T_3}_{SO(10)}=-2A^{R_3}_{SO(10)}, \qquad 
A^{T_3}_{SU(4)}=-2A^{R_3}_{SU(4)}, \nonumber \\
& & A^{T_3}_{SO(12)}=-2A^{R_3}_{SO(12)}, \qquad 
A^{T_3}_{SU(2)}=-2A^{R_3}_{SU(2)}, 
\end{eqnarray}
because this model has only massless modes with $r_3=0,1/2,1$.
Indeed, all of $Z_4$ orbifold models include only 
massless modes with $r_3=0,1/2,1$.
Furthermore, all of $Z_N$ orbifold models with $v_i=1/2$ 
have only massless modes with $r_i=0,1/2,1$.
Thus, the above relation (\ref{relation:T-R}) 
holds true in such $Z_N$ orbifold models.
That is also true for $R_1$-anomalies in $Z_2 \times Z_M$ 
orbifold models with $v_1=(1/2,-1/2,0)$ and $v_2=(0,v_2,-v_2)$.

Such relation between T-duality anomalies and 
R-anomalies (\ref{relation:T-R}) would be important, 
because the GS condition on R-anomalies leads to a certain 
condition on the T-duality anomalies even including the 
N=2 subsector.
For example, in the above $Z_4$ orbifold model, 
the following condition is required 
\begin{equation}
A^{T_3}_{SO(10)}=A^{T_3}_{SU(4)}=
A^{T_3}_{SO(12)}=A^{T_3}_{SU(2)} \qquad ({\rm mod}~~2).
\end{equation}

\section{Phenomenological implications}

\subsection{Symmetry breaking of
the discrete R-symmetries}

\subsubsection{Nonperturbative breaking}

If the discrete R-symmetries are anomalous, they
  are broken by nonperturbative effects  at  low energy.
This is because,  for
the GS mechanism to take place,
  the axionic part of the dilaton $S$ should transform
non-linearly under the anomalous symmetry.
This means that a term like $e^{-aS}$
with a constant $a$ has a
definite charge $R_i^S$ under the anomalous symmetry.
Nonperturbative effects can therefore induce
terms like $e^{-aS}\Phi^1 \cdots \Phi^n$
with matter fields $\Phi^a$, where the total
charge satisfies the condition for allowed couplings, 
i.e.  $R^S_i+\sum_a R^a_{i}=1$ (mod $N_i$).
This implies that  below the scale of the 
vacuum expectation value (VEV) of $S$,
  such non-invariant terms can
  appear in a low-energy effective Lagrangian.
  The canonical dimension of the non-invariant operator
$e^{-aS}\Phi^1 \cdots \Phi^n$  that can be generated by the
nonperturbative effects depends of course on the R charge $R^S$.
If the smallest dimension is lager than four, they will be
suppressed
by certain powers of the string scale.
However,  the operator can produce
non-invariant mass terms like $m \Phi \Phi'$,
because some of the chiral superfields may acquire VEVs.
One should worry about such cases.
Needless to say that small higher dimensional  terms  would be useful
in phenomenological applications such as explaining
fermion masses.

In the case that
the smallest dimension is smaller than three,
the anomalous discrete R symmetry
has less power to constrain the low-energy theory.

\subsubsection{Spontaneous breaking}

In the discussion above, we have considered
R-symmetry breaking by nonperturbative effects when
R-symmetries are anomalous.
Here we comment on another type of symmetry breaking;
they can be broken spontaneously by the VEVs of scalar fields in the form
$U(1)\times R \rightarrow R'$.
That is, we consider a spontaneous symmetry breaking, where
some scalar fields with non-vanishing $U(1)$ and $R$ charges
develop their VEVs and they break
$U(1)$ and $R$ symmetries in such a way that  an unbroken $R'$ symmetry
remains intact.
(Its order is denoted by $N'$ below.)
Even in such symmetry breaking, we can obtain
the GS condition for the unbroken $R'$ from the GS condition for the $U(1)$
and R-anomalies.
Suppose that we have the GS condition for the $U(1)$ symmetry as
\begin{equation}
Tr Q T({\bf R}_{G_a})/k_a= Tr Q T({\bf R}_{G_b})/k_b,
\end{equation}
where $Q$ is the $U(1)$ charge.
Since the unbroken $R'$ charge is a linear combination of $R_i$ and  $Q$,
 the mixed anomalies for $R'$ should also satisfy the
GS condition,
\begin{equation}
Tr R' T({\bf R}_{G_a})/k_a= Tr R' T({\bf R}_{G_b})/k_b.
\end{equation}
Here the anomaly coefficients $Tr R' T({\bf R}_{G_a})$ are
defined modulo $N'T({\bf R}^{(f)}_{G_a})$.

Through the symmetry breaking $U(1)\times R \rightarrow R'$,
some matter fields may  gain mass terms like
\begin{equation}
W\sim m \Phi \bar \Phi.
\end{equation}
Such a pair of the matter fields $\Phi$ and $\bar \Phi$ should form
 a vector-like representation of $G_{a}$ and have
opposite $R'$ charges of the unbroken $R'$ symmetry.
The heavy modes of this type have therefore no contribution to the 
mixed anomalies between the gauge symmetry $G_a$ and
the unbroken $R'$ symmetry.
This implies that  the above GS condition for the unbroken $R'$ remains
unchanged even after the spontaneous symmetry breaking.
The symmetry breaking
$U(1)\times R \rightarrow R'$ also allows
 Majorana mass terms like
\begin{equation}
W\sim m \Phi \Phi.
\end{equation}
This type of Majorana mass terms can appear for an even
 order $N'$ of the $R'$ symmetry if the $R'$
charge of $\Phi$ is $N'/2$ and $\Phi$ is
in a real representation  ${\bf R}_{G_a}$
of the unbroken gauge group $G_a$.
The field $\Phi$ contributes to the anomaly coefficient as
$\frac{N'}{2}T({\bf R}_{G_a})$.
That however may change only the modulo-structure of the anomaly
coefficients.
For $SU(N)$ gauge group, this contribution is
obtained as $\frac{N'}{2}\times
({\rm integer})$.
Thus, the modulo-structure does not change, that is,
the anomaly coefficients  $Tr R' T({\bf R}_{G_a})$ are defined
modulo $N'/2$.
However, for other gauge groups, the modulo-structure
of the anomaly coefficients may change.

\subsection{
Gravity-induced supersymmetry breaking and
  Gauge symmetry breaking}
The most important difference of the discrete R-symmetries
compared with T-duality in phenomenological applications
comes from the fact that (for the heterotic orbifold
string models) the moduli and  dilaton superfields have vanishing 
R-charges.
The VEVs of their bosonic components do not therefore violate
the discrete R-symmetries in  the  perturbation theory.
(We have discussed above the nonperturbative effects due to the VEV of 
the dilaton, which may be small in a wide class of  models.)
However,
the F-components of the moduli and  dilaton superfields
have non-zero R-charges. Therefore, since the VEVs of these
F-components generate soft-supersymmetry breaking (SSB) terms
at low energy, the SSB terms do not have to respect the discrete 
R-symmetries. \footnote{
Whether
  the nonperturbative effects due to the VEV of the dilaton
  do play an important roll in the SSB sector depends on the
  R charge of the dilaton, and one has to check it explicitly
  for a given model.}
Fortunately,  in the visible sector,
the scale of the R-symmetry breaking must be of the same
order as that of  supersymmetry breaking. 
 If the order of the discrete R-symmetry is even, 
the VEVs of these F-components break the discrete R-symmetry
down to its subgroup $Z_2$, an R-parity.
That is  an interesting observation because it may be an origin
of the R-parity of the minimal supersymmetric standard model (MSSM).

Gauge symmetry breaking can be achieved by VEVs of chiral
supermultiplets in a non-trivial representation of the gauge group
or by non-trivial Wilson lines. Clearly, if the chiral supermultiplets
have  vanishing R-charges and only their scalar
components acquire VEVs,  the discrete  R-symmetries remain
unbroken. Similarly, the Wilson lines do not break the discrete R-symmetries
because gauge fields have no R charge.
As a consequence, the discrete R-symmetries have a good chance
to be intact at low energy if the nonperturbative effects are small.

\subsection{Constraints on low-energy beta-functions}
Only anomaly-free discrete R-symmetries
remain as intact symmetries in a low-energy effective theory.
Obviously, the model with anomaly-free
discrete R-symmetries corresponds to $A^{R_i}_{G_a}=0$
(mod $N_iT({\bf R}^{(f)}_{G_a}))$.
Consider  for instance $SU(N)$ gauge groups for which
$T({\bf R}^{(f)}_{G_a})=1/2$ is usually satisfied.
Then in models, which have
no oscillator mode in a non-trivial representations of $SU(N)$,
  the relation between R-anomalies
and beta-function coefficients lead to
\begin{equation}
b_a = 2 A^{}_{G_a}=0,
\end{equation}
mod $N_i$ for any gauge group $G_a$.
For example, the $Z_3$ orbifold model
with anomaly-free R-symmetries leads to
$b_a=3n_a$ with integer $n_a$, while
the $Z_4$ orbifold model with anomaly-free R-symmetries
leads to $b_a=2n_a$.
Similarly,
$b_a=1$ would be possible in $Z_6$-II orbifold models
because $N_i=(6,3,2)$ as one can see from Table 1.

Even for anomalous discrete R-symmetries,
the GS condition for R-anomalies and the relation between
beta-function coefficients (\ref{GS-R}), (\ref{anomR-b}), 
(\ref{GS-b}) would have phenomenological
implications. As discussed at the beginning in this section,
the non-perturbative effects  can generate
operators like $e^{-aS}\Phi^1 \cdots \Phi^n$.
If its canonical dimension is larger than four, its contribution
to low-energy beta-functions may be assumed to be small.
\footnote{If the operator produces
non-invariant mass terms like $M \Phi \Phi'$ with $M$
larger than the low-energy scale,
the low-energy spectrum may change. Then
the power of the discrete R-symmetries decreases.}

As for the MSSM
we find $b_3=-3$ and $b_2=1$ for $SU(3)$ and $SU(2)$, respectively.
That is, we have $b_2 - b_3=4$, implying
the MSSM can not be realized, e.g. in $Z_3$ orbifold models,
because $Z_3$ orbifold models require
$b_a - b_b=0$ mod $3$ if the effects of the symmetry breaking
of the discrete R-symmetries can be neglected.
Similarly, the model with $b_2 - b_3=4$ can not be
obtained in the $Z_6$-I, $Z_7$ or $Z_{12}$-I orbifold models.

\section{Conclusion}

We have studied anomalies of the discrete R-symmetries
in heterotic orbifold models.
They are remnants of $SU(4)_R$ symmetry which,
along with  extended $N=4$ supersymmetry,
is explicitly broken by orbifolding. 
We have found that 
the mixed anomalies for different gauge groups  
satisfy the universal GS condition.
Therefore, 
these anomalies 
can be  canceled by the GS mechanism,
which remains to be proven at the string theory level.
As a byproduct,
we have found a relation between 
the anomaly coefficients
of the discrete R-symmetries 
and one-loop beta-function coefficients.
In particular, in the case that
the contribution coming from the oscillator modes for
the chiral matter fields in 
non-trivial representations of a gauge group
vanishes, the anomaly coefficient corresponding to
the sum of the discrete R-symmetry anomaly
is exactly proportional  to the  one-loop beta-function
coefficient of the corresponding gauge coupling.

In a wide class of models,  the discrete R-symmetries 
may be unbroken at low energy.
The main reason for this is that the moduli 
superfields have  vanishing R-charges.
This should be contrasted to the case of T-duality,
where the moduli fields transform non-trivially
under the T-duality transformation.
We have  studied the relation between anomalies of
the discrete R-symmetries and T-duality.
We have argued that the discrete R-symmetries 
have a good chance to be unbroken down to
the supersymmetry breaking scale.
Even below this scale a $Z_2$ subgroup 
is  unbroken, which  may be an origin of 
the R-parity of the MSSM.
In fact, the R-parity of the MSSM is completely anomaly-free,
indicating that it has a stringy origin.

Our investigation
on the discrete R-symmetries in heterotic orbifold models 
could be extended to other types of heterotic models, e.g. 
free fermionic construction \cite{Antoniadis:1989zy} and 
Gepner models \cite{Gepner:1987vz} as well as 
Calabi-Yau models.
Furthermore, our studies 
can be extended to type IIA and IIB string theories 
with D-branes, e.g. intersecting/magnetized D-brane models.
This however would be beyond the scope of the present paper,
and we will leave it   to our future study.
At last we emphasize that
string models have other discrete symmetries.
For example, heterotic orbifold models have non-abelian discrete flavor 
symmetries \cite{Kobayashi:2006wq}.
They may be identified with the non-abelian discrete flavor 
symmetries which have been recently  introduced
in constructing  low-energy flavor models \cite{Altarelli:2007cd}.
Further investigations in this direction are certainly
necessary to link the non-abelian discrete flavor  symmetries 
from the top and the bottom with each other.

\subsection*{Acknowledgement}
K.~S.~C. \/ is supported in part by the European Union 6th framework program 
MRTN-CT-2004-503069 "Quest for unification",
MRTN-CT-2004-005104 "ForcesUniverse", MRTN-CT-2006-035863 
"UniverseNet and SFB-Transregio 33 "The Dark Univeres"  
by Deutsche Forschungsgemeinschaft (DFG).
T.~K.\/ and J.~K.\/ are supported in part by the Grand-in-Aid for 
Scientific Research \#1754025,\#18540257 and \#19034003, respectively. 
T.~K.\/ is also supported in part by the Grant-in-Aid for the 21st Century 
COE ``The Center for Diversity and Universality in Physics'' from the 
Ministry of Education, Culture, Sports, Science and Technology of Japan.


\begin{thebibliography}{99}

\bibitem{Altarelli:2007cd}
See, e.g. , \\
G.~ Altarelli,
arXiv:0705.0860 [hep-ph];
E.~Ma,
 arXiv:0705.0327 [hep-ph]
 and references therein.
 
 \bibitem{dbkaplan}
D.~B.~Kaplan and M.~Schmaltz,
Phys. Rev. D {\bf 49}, 3741 (1994);
L.J.~Hall and H.~Murayama,
Phys. Rev. Lett. {\bf 75}, 3985  (1995)  ;
C.D.~Carone, L.J.~Hall and H.~Murayama,
Phys. Rev. D {\bf 53}, 6282 (1996).

\bibitem{babu}
K.S.~Babu, T.~Kobayashi and J.~Kubo, 
Phys. Rev. D {\bf 67}, 075018 (2003);
%
K.~Hamaguchi, M.~Kakizaki and M.~Yamaguchi, 
Phys. Rev. D {\bf 68}, 056007 (2003);
T.~Kobayashi, J.~Kubo and H.~Terao, 
Phys. Lett. B {\bf 568}, 83 (2003);
%
 G. G. Ross, L.~Velasco-Sevilla
 and  Oscar Vives, Nucl. Phys. B {\bf 692}, 50 (2004);
S. F.~King and G. G.~Ross, Phys. Lett. B {\bf 520}, 243 (2001);
B{\bf 574,} 239 (2003);
G. G.~Ross and L.~Velasco-Sevilla, Nucl. Phys. B {\bf 653}, 3 (2003);
Ki-Y.~Choi, Y.~Kajiyama,
J.~Kubo and H.M.~Lee,  Phys. Rev. D {\bf 70}, 055004 (2004);
%
N.~Maekawa and T.~Yamashita,
JHEP {\bf  0407,} 009 (2004);
K.S.~Babu and J.~Kubo, Phys. Rev. D {\bf 71}, 056006 (2005);
%
T.~Yamashita, hep-ph/0503265;
I.~de Medeiros Varzielas and  G.G.~Ross, hep-ph/0612220;
%
  P.~Ko, T.~Kobayashi, J.~h.~Park and S.~Raby,
  arXiv:0704.2807 [hep-ph].




\bibitem{murayama}
  L.~E.~Ibanez and G.~G.~Ross,
  Nucl.\ Phys.\  B {\bf 368}, 3 (1992);
%
  S.~P.~Martin,
  Phys.\ Rev.\  D {\bf 46}, 2769 (1992);
%
%
H.~Murayama and D.B.~Kaplan,  
Phys.~Lett. B {\bf 336},  221 (1994);
V.~Ben-Hamo and Y.~Nir,
Phys.~Lett.~B {\bf 339}, 77 (1994);
C.D.~Carone, L.J.~Hall and H.~Murayama,
Phys. Rev. D {\bf 53}, 6282 (1996).
\bibitem{kakizaki}  
M.~Kakizaki and M.~Yamaguchi,
JHEP~{\bf 0206,} 032 (2002);
R.~Harnik,~D.T.~Larson,~H.~Murayama and M.~Thormeier,
Nucl.~Phys.~B {\bf 706}, 372 (2005);
E.~Itou, Y.~Kajiyama and J.~Kubo,
Nucl.~Phys.~B {\bf 743}, 74 (2006).

 
\bibitem{Dine:1988kq}
  M.~Dine and N.~Seiberg,
  Nucl.\ Phys.\  B {\bf 306}, 137 (1988).



\bibitem{Green:1984sg}
  M.~B.~Green and J.~H.~Schwarz,
  Phys.\ Lett.\  B {\bf 149}, 117 (1984).

\bibitem{Witten:1984dg}
  E.~Witten,
  Phys.\ Lett.\  B {\bf 149}, 351 (1984);
%
%
  M.~Dine, N.~Seiberg and E.~Witten,
  Nucl.\ Phys.\  B {\bf 289}, 589 (1987);
%
%
  W.~Lerche, B.~E.~W.~Nilsson and A.~N.~Schellekens,
  Nucl.\ Phys.\  B {\bf 289}, 609 (1987);
  J.~J.~Atick, L.~J.~Dixon and A.~Sen,
  Nucl.\ Phys.\  B {\bf 292}, 109 (1987);
%
  M.~Dine, I.~Ichinose and N.~Seiberg,
  Nucl.\ Phys.\  B {\bf 293}, 253 (1987).


\bibitem{Ibanez:1998qp}
  L.~E.~Ibanez, R.~Rabadan and A.~M.~Uranga,
  Nucl.\ Phys.\  B {\bf 542}, 112 (1999);
%
  Z.~Lalak, S.~Lavignac and H.~P.~Nilles,
  Nucl.\ Phys.\  B {\bf 559}, 48 (1999).





\bibitem{Schellekens:1986xh}
  A.~N.~Schellekens and N.~P.~Warner,
  Nucl.\ Phys.\  B {\bf 287}, 317 (1987).

\bibitem{Kobayashi:1996pb}
  T.~Kobayashi and H.~Nakano,
  Nucl.\ Phys.\  B {\bf 496}, 103 (1997).



\bibitem{Derendinger:1991hq}
  J.~P.~Derendinger, S.~Ferrara, C.~Kounnas and F.~Zwirner,
  Nucl.\ Phys.\  B {\bf 372}, 145 (1992).

\bibitem{Ibanez:1992hc}
  L.~E.~Ibanez and D.~L\"ust,
  Nucl.\ Phys.\  B {\bf 382}, 305 (1992).

\bibitem{Ibanez:1991zv}
  L.~E.~Ibanez, D.~L\"ust and G.~G.~Ross,
  Phys.\ Lett.\  B {\bf 272}, 251 (1991).




\bibitem{Kawabe:1993pz}
  H.~Kawabe, T.~Kobayashi and N.~Ohtsubo,
  Phys.\ Lett.\  B {\bf 325}, 77 (1994);
%
  Nucl.\ Phys.\  B {\bf 434}, 210 (1995).


\bibitem{Ibanez:1991hv}
  L.~E.~Ibanez and G.~G.~Ross,
  Phys.\ Lett.\  B {\bf 260} (1991) 291;
  T.~Banks and M.~Dine,
  Phys.\ Rev.\  D {\bf 45}, 1424 (1992);
  L.~E.~Ibanez,
  Nucl.\ Phys.\  B {\bf 398}, 301 (1993);
  K.~Kurosawa, N.~Maru and T.~Yanagida,
   Phys. Lett. B{\bf 512}, 203 (2001);
  J.~Kubo and D.~Suematsu, 
  Phys. Rev. D{\bf 64}, 115014 (2001);
K. S.~Babu, I.~Gogoladze and K.~Wang, 
  Nucl. Phys. B{\bf 660}, 332 (2003);
  M.~Dine and M.~Graesser,
  JHEP {\bf 0501}, 038 (2005);
  T.~Araki,
  arXiv:hep-ph/0612306;
H.~Dreiner and M.~Thormeier, 
Phys. Rev. D{\bf 69}, 053002 (2004);
  H.~Dreiner, H.~Murayama and M.~Thormeier,
     Nucl. Phys. B{\bf 729}, 278 (2005);
  A.~H.~Chamseddine and H.~K.~Dreiner,
  Nucl.\ Phys.\  B {\bf 458}, 65 (1996);
  H.~K.~Dreiner, C.~Luhn and M.~Thormeier,
  Phys.\ Rev.\  D {\bf 73}, 075007 (2006);
  H.~K.~Dreiner, C.~Luhn, H.~Murayama and M.~Thormeier,
  arXiv:hep-ph/0610026.







\bibitem{Dixon}
L.~J.~Dixon, J.~A.~Harvey, C.~Vafa and E.~Witten, Nucl.\ Phys.\ B\
{\bf 261}, 678 (1985);
Nucl.\ Phys.\ B\ {\bf 274}, 285 (1986).



\bibitem{IMNQ}
L.~E.~Ib\'a\~nez, H.-P.~Nilles and F.~Quevedo, Phys.\ Lett.\ B\
{\bf 187}, 25 (1987);
L.~E.~Ib\'a\~{n}ez, J.~E.~Kim, H.-P.~Nilles and F.~Quevedo, Phys.\
Lett.\ B\ {\bf 191}, 282 (1987);
L.~E.~Ib\'a\~{n}ez, J.~Mas,
H.~P.~Nilles and F.~Quevedo, Nucl.\ Phys.\ B\ {\bf 301}, 157
(1988);
A.~Font, L.~E.~Ib\'a\~{n}ez, F.~Quevedo and A.~Sierra,
Nucl.\ Phys.\ B\ {\bf 331}, 421 (1990);
  A.~Font, L.~E.~Ibanez, H.~P.~Nilles and F.~Quevedo,
  Phys.\ Lett.\  {\bf 210B}, 101 (1988)
  [Erratum-ibid.\  B {\bf 213}, 564 (1988)];
  Phys.\ Lett.\  B {\bf 213}, 274 (1988);
  J.~A.~Casas and C.~Munoz,
  Phys.\ Lett.\  B {\bf 209}, 214 (1988);
  J.~A.~Casas and C.~Munoz,
  Phys.\ Lett.\  B {\bf 214}, 63 (1988);
%
D.~Bailin, A.~Love and
S.~Thomas, Phys.\ Lett.\ B\ {\bf 194}, 385 (1987);
Y.~Katsuki, Y.~Kawamura, T.~Kobayashi, N.~Ohtsubo, Y. Ono and
K.~Tanioka, Nucl.\ Phys.\ B\ {\bf 341}, 611 (1990).



\bibitem{Kobayashi:2004ud}
T.~Kobayashi, S.~Raby and R.~J.~Zhang,
Phys.\ Lett.\ B {\bf 593}, 262 (2004);
%
%
Nucl.\ Phys.\ B {\bf 704}, 3 (2005).

\bibitem{Forste:2004ie}
S.~Forste, H.~P.~Nilles, P.~K.~S.~Vaudrevange and A.~Wingerter,
Phys.\ Rev.\ D {\bf 70}, 106008 (2004);
%
W.~Buchmuller, K.~Hamaguchi, O.~Lebedev and M.~Ratz,
 Nucl.\ Phys.\ B {\bf 712}, 139 (2005);
%
  Phys.\ Rev.\ Lett.\  {\bf 96}, 121602 (2006);
  %
  arXiv:hep-th/0606187;
%
%
  S.~Forste, H.~P.~Nilles and A.~Wingerter,
  Phys.\ Rev.\ D {\bf 72}, 026001 (2005);
%
  Phys.\ Rev.\ D {\bf 73}, 066011 (2006);
%
  K.~S.~Choi, S.~Groot Nibbelink and M.~Trapletti,
  JHEP {\bf 0412} (2004) 063;
%
%
%
  H.~P.~Nilles, S.~Ramos-Sanchez, P.~K.~S.~Vaudrevange and A.~Wingerter,
  JHEP {\bf 0604}, 050 (2006);
  J.~E.~Kim and B.~Kyae,
  arXiv:hep-th/0608085;
%
  arXiv:hep-th/0608086.
%


\bibitem{Choi-Km}
K.~S.~Choi and J.~E.~Kim, ``Quarks and Leptons from Orbifolded 
Superstring'', Springer (2006).

\bibitem{Ibanez:1992uh}
  L.~E.~Ibanez and D.~L\"ust,
  Phys.\ Lett.\  B {\bf 302}, 38 (1993).



\bibitem{Kobayashi:2006wq}
  T.~Kobayashi, H.~P.~Nilles, F.~Ploger, S.~Raby and M.~Ratz,
  Nucl.\ Phys.\  B {\bf 768}, 135 (2007).




\bibitem{Friedan:1985ge}
  D.~Friedan, E.~J.~Martinec and S.~H.~Shenker,
  Nucl.\ Phys.\ B {\bf 271}, 93 (1986).

\bibitem{Dixon:1986qv}
  L.~J.~Dixon, D.~Friedan, E.~J.~Martinec and S.~H.~Shenker,
  Nucl.\ Phys.\ B {\bf 282}, 13 (1987).


\bibitem{Font:1988mk}
  A.~Font, L.~E.~Ib\'{a}\~{n}ez and F.~Quevedo,
  Phys.\ Lett.\ B {\bf 217} (1989) 272.

\bibitem{Faraggi:2006bs}
  A.~E.~Faraggi, S.~Forste and C.~Timirgaziu,
  JHEP {\bf 0608}, 057 (2006);
%
  S.~Forste, T.~Kobayashi, H.~Ohki and K.~j.~Takahashi,
  JHEP {\bf 0703}, 011 (2007);
%
  K.~j.~Takahashi,
  arXiv:hep-th/0702025;
%
  F.~Ploger, S.~Ramos-Sanchez, M.~Ratz and P.~K.~S.~Vaudrevange,
  arXiv:hep-th/0702176.


\bibitem{Kobayashi:1991rp}
  T.~Kobayashi and N.~Ohtsubo,
  Int.\ J.\ Mod.\ Phys.\  A {\bf 9}, 87 (1994).


\bibitem{Katsuki:1989cs}
  Y.~Katsuki, Y.~Kawamura, T.~Kobayashi, N.~Ohtsubo, Y.~Ono and K.~Tanioka,
  DPKU-8904.




\bibitem{jones}
  D.R.T~Jones and L.~Mezincescu,
  Phys. Lett. B {\bf 136}, 242 (1984);
  B {\bf 138}, 293 (1984);
   P.~West, Phys. Lett. B {\bf 137}, 371 (1984);
A.J.~Parkes and P.C~West, Phys. Lett. B {\bf 138}, 99 (1984);
Nucl. Phys. B {\bf 256}, 340 (1985);
A.J.~Parkes,
  Phys. Lett. B {\bf 156}, 73 (1985);
D.R.T~Jones and A.J.~Parkes,
  Phys. Lett. B {\bf 160}, 267 (1985).

  \bibitem{piguet1}
O.~Piguet and K.~Sibold, Int. Mod. Phys. A {\bf 1},
913 (1986);
Phys. Lett. B {\bf 177}, 373 (1986).

\bibitem{ferrara}
S.~Ferrara and B.~Zumino, Nucl. Phys. B {\bf 87}, 207 (1975).

\bibitem{piguet2}
O.~Piguet and K.~Sibold,
Nucl. Phys. B {\bf 196}, 428 (1982); B {\bf 196}, 447 (1982).








\bibitem{Dixon:1990pc}
  L.~J.~Dixon, V.~Kaplunovsky and J.~Louis,
  Nucl.\ Phys.\  B {\bf 355}, 649 (1991);

  I.~Antoniadis, K.~S.~Narain and T.~R.~Taylor,
  Phys.\ Lett.\  B {\bf 267}, 37 (1991).


\bibitem{Antoniadis:1989zy}
  H.~Kawai, D.~C.~Lewellen and S.~H.~H.~Tye,
  Phys.\ Rev.\ Lett.\  {\bf 57} (1986) 1832
  [Erratum-ibid.\  {\bf 58} (1987) 429],
  Nucl.\ Phys.\ B {\bf 288} (1987) 1;
  I.~Antoniadis, C.~P.~Bachas and C.~Kounnas,
  Nucl.\ Phys.\ B {\bf 289} (1987) 87;
%
  I.~Antoniadis, J.~R.~Ellis, J.~S.~Hagelin and D.~V.~Nanopoulos,
  Phys.\ Lett.\  B {\bf 205}, 459 (1988);
%
  Phys.\ Lett.\  B {\bf 208}, 209 (1988)
  [Addendum-ibid.\  B {\bf 213}, 562 (1988)];
  A.~E.~Faraggi, D.~V.~Nanopoulos and K.~j.~Yuan,
  Nucl.\ Phys.\  B {\bf 335}, 347 (1990).


\bibitem{Gepner:1987vz}
  D.~Gepner,
  Phys.\ Lett.\ B {\bf 199}, 380 (1987);
  Nucl.\ Phys.\ B {\bf 296}, 757 (1988).


\end{thebibliography}
\end{document}